# Inductively Coupled Plasma Driven by Asymmetric Triangular Current Waveform

Shahid Rauf,[1,a)] Tianhong Wang,[1] Jason Kenney,[1] Dmytro Sydorenko,[2] and Igor D. Kaganovich[3]

[1]Applied Materials, Inc., Santa Clara, CA 95054, USA

[2]University of Alberta, Edmonton, Alberta T6G2E1, Canada

[3]Princeton Plasma Physics Laboratory, Princeton, New Jersey 08543, USA

[a)]Author to whom correspondence should be addressed: shahid_rauf@amat.com

## Abstract

Two-dimensional particle-in-cell simulations of an inductively coupled plasma (ICP) are used to investigate the influence of radio-frequency (RF) current waveform and frequency on plasma characteristics, collision processes, and the electron velocity distribution function (EVDF). Plasmas driven by sinusoidal coil current over the frequency range 1–30 MHz and by asymmetric triangular waveform current at 1 MHz are examined at 25 mTorr. For sinusoidal excitation, significantly lower coil current is required at higher frequency to achieve comparable electron density, consistent with the scaling of inductive electric field with frequency. Plasma production shifts toward the plasma – dielectric interface with increasing frequency due to reduced skin depth. Despite differences in power deposition localization, strong diffusion in the small chamber results in broadly similar plasma density profiles across frequencies. For asymmetric triangular waveform current, the inductive electric field is strongly enhanced during the faster current ramp-down phase, leading to pronounced temporal modulation of electron heating. Increasing waveform duty cycle (DC) enhances the contribution of higher current harmonics, which results in higher steady-state electron density. Excitation and ionization processes also shift closer to the plasma –dielectric interface. The EVDF becomes increasingly asymmetric with increasing DC, reflecting preferential electron acceleration during the current ramp-down phase. Time-resolved analysis shows that excitation and ionization rates peak sharply during this phase, with ionization exhibiting stronger relative enhancement. The intense plasma current produced during the current ramp-down phase persists beyond the reversal of the source electric field due to electron inertia and collisional effects. Consequently, inductive electron power deposition continues in the plasma for several 10s of ns after the coil current reverses direction. These results demonstrate that RF current waveform asymmetry provides a viable mechanism for controlling electron power deposition, EVDF, and electron-impact collisional processes in ICPs.

## 1. Introduction

Inductively coupled plasmas (ICP) are widely employed in microelectronics fabrication. As microelectronics device dimensions scale down to a few nm, increasingly precise control over plasma properties is required. Pulsed plasma operation, introduced more than twenty years ago, enabled control over ion and radical fluxes and ion energy on milliseconds timescale. [1] Radio frequency (RF) power pulsing provides a means to regulate the relative duration of key fundamental processes on the wafer surface, including passivation, polymerization, etching, and byproduct removal. Modern ICP systems employ multiple RF power sources for plasma generation and ion energy control. Consequently, increasingly sophisticated pulsing schemes have been developed over time. [2] In recent years, significant emphasis has been placed on ion energy distribution function (IEDF) control. Waveform tailoring concepts [3, 4] employing non-sinusoidal bias voltage waveforms have been used to control the shape of the IEDF and tune the relative populations of low and high energy ions. In contrast, the electron energy distribution function (EEDF), which governs the relative population of electrons at different energies, has received less attention. EEDF determines the relative rate of different electron impact processes and, hence, controls the population of chemically reactive species available for surface processing. Better control over EEDF would allow optimizing the plasma chemistry in ICPs to improve the profile during reactive ion etch (by balancing sidewall and feature bottom passivation) and the properties of deposited films. This paper investigates whether the EEDF in ICP discharges can be tailored using non-sinusoidal current waveforms. A particle-in-cell based kinetic plasma model is used for this investigation to accurately examine EEDF evolution during the RF cycle.

The literature on ICP plasmas is extensive. Therefore, only studies relevant to the present work are reviewed here. EEDF in ICPs has been investigated both experimentally and theoretically. Godyak and Kolobov [5] reported EEDF measurements in low-pressure ICP at several driving frequencies. They attributed the observed frequency dependence of the electron energy probability function (EEPF) to collisionless electron heating. Godyak, Piejak, and Alexandrovich [6] published additional measurements in ICP over a range of frequencies, powers, and pressures. They demonstrated that the EEPF is Maxwellian at high pressures (100 and 300 mTorr) but transitions to a three-temperature distribution at low-pressures (below 10 mTorr). Vasenkov and Kushner [7] used an electron Monte Carlo model to study the EEDF in low pressure ICPs. They showed that electron-electron collisions can deplete low energy electrons in the EEDF, particularly in regimes dominated by collisionless electron heating.

The influence of driving frequency on ICP characteristics and kinetics has been widely studied. Kolobov and Godyak [8] examined low frequency ICP operation, particularly in systems employing ferromagnetic cores where inductive magnetic field penetration in plasma is limited. Yang *et al.* [22] demonstrates that adding low-frequency 2 MHz power to a 13.56 MHz ICP can significantly improve radial plasma and etch rate uniformity by enhancing the central plasma density. Terentev *et al.* [9] used a fluid model of ICP at 113 Pa to investigate the effect of frequency on plasma characteristics. They found that several plasma properties including electron density have a non-monotonic dependence on frequency. Rauner and coauthors [10] examined the influence of frequency on RF power transfer efficiency in a low pressure $H_2$ ICP and reported better power transfer efficiency at higher frequency. Hao *et al.* [11] investigated the effect of frequency (2 vs. 13.56 MHz) on a cylindrical ICP. They reported higher electron density at 13.56 MHz compared to 2 MHz for 0.8 – 1.2 Pa (6 – 9 mTorr) at a given power, which was attributed to better electron heating efficiency at 13.56 MHz. Jun and Chang [12] described a 40 MHz ICP and observed lower electron densities compared to 13.56 MHz at the same power based on Langmuir probe measurements. Gao *et al.* [13] compared plasma characteristics in 2 and 13.56 MHz ICPs and found the electron density to be comparable at these frequencies. Si *et al.* [14] used a fluid ICP model to study current harmonics at different frequencies. They showed that nonlinear Lorenz force strongly influences harmonic generation, particularly at low frequencies. Xu *et al.* [15] described an ICP at 500 kHz and demonstrated that it operates efficiently in both the E and H modes.

Due to high plasma densities in ICPs, fully kinetic multi-dimensional simulations of ICPs remain uncommon. Kolobov and Godyak [16] used the non-local approach to compute the EEDF in ICPs. Their computations matched well with experiments in the 0.9 – 4.5 Pa (6.75 – 33.75 mTorr) pressure range. Ramamurthi *et al.* [17] described a self-consistent 1D ICP model with a non-local electron kinetics module. Detailed comparison of the kinetic model with a local model as well as comparison of the computed EEDF with experiments were presented. Logue and Kushner [18] developed a hybrid ICP model with an electron Monte Carlo model for computing the EEDF. Using simulation of pulsed Ar and $N_2$ ICPs, they explained how the duty cycle impacts the EEDF and the source function of various reactive species. Nishida, Mattei, and colleagues [19–21] described a fully implicit PIC model of $H_2$ ICP in cylindrical geometry. Their model captured the E-to-H transition and plasma dynamics in both modes of operation. Modeling results agreed qualitatively with optical emission measurements. Chen *et al.* [23, 24] developed a 2-dimensional (2D) cylindrical geometry PIC model for ICPs. They examined the impact of phase between ICP and RF bias sources and found it to have major impact on the plasma characteristics in the E mode. They also examined power deposition in the E and H modes. In recent work, Chen *et al.* [25] proposed coil current waveform tailoring as a new method for EEPF and plasma chemistry control in ICPs. Using four harmonics to produce sawtooth-like current waveforms, they demonstrated control over the EEPF and the ionization-to-excitation ratio. Takao *et al.* [26, 27] presented a 2D model of a miniature ICP discharge that combined a frequency domain electromagnetic model with an electrostatic PIC model. While the spatial plasma distribution was similar for the collisional and kinetic models, significant differences in EEDF were observed. Fu *et al.* [28] described an implicit electromagnetic PIC model, which was used to understand the effect of RF coil current, driving frequency, and gas pressure on Xe plasma characteristics. Main *et al.* [29] discussed an implicit PIC framework for ICPs focusing on ion energy and angular distributions.

This article is organized in the following manner. The computational model is described in Sec. 2. Modeling results are presented in Sec. 3, and our concluding remarks are in Sec. 4.

## 2. Computational Model

The simulations in this article have been done using the 2D particle-in-cell (PIC) modeling code developed by Sydrenko *et al.* [30] This model uses the Darwin approximation for electromagnetic fields, which self-consistently captures low-frequency electromagnetic phenomena while eliminating the propagation of light waves. As the model uses the direct implicit electrostatic solver, time steps can be significantly longer than those required by fully electromagnetic explicit PIC methods, making it well suited for simulating ICP discharges. The model resolves two spatial dimensions and three velocity components for charged particles. Electrons and ions are represented as macro-particles, and particle motion is coupled self-consistently to the electromagnetic fields through charge and current deposition onto staggered Cartesian grids.

The electric field $\mathbf{E}$ is decomposed into irrotational (electrostatic) $\mathbf{E}_{irr}$ and solenoidal (inductive) $\mathbf{E}_{sol}$ components: $\mathbf{E} = \mathbf{E}_{irr} + \mathbf{E}_{sol}$. The irrotational field is obtained from a direct implicit formulation of Poisson's equation, which allows grid spacing much larger than the Debye length. The solenoidal electric field $\mathbf{E}_{sol}$ is computed using the Darwin formulation:

$$\nabla^2 \mathbf{E}_{sol} = \mu_0 \left( \frac{\partial \mathbf{J}_{sol(plasma)}}{\partial t} + \frac{\partial \mathbf{J}_{ext}}{\partial t} \right) \quad (1)$$

where $\mathbf{J}_{sol(plasma)}$ is the solenoidal part of electric current in the plasma and $\mathbf{J}_{ext}$ is the external divergence-free current in the coils. In this article, $\mathbf{J}_{ext} = J_{z(ant)} \hat{\mathbf{z}}$ and the $z$-component of the electric field is referred to as the inductive electric field. The Darwin approximation neglects the transverse displacement current associated with electromagnetic radiation. The magnetic field is computed self-consistently from the solenoidal current using a vector-potential formulation in the Coulomb gauge, ensuring divergence-free magnetic fields and robust energy behavior. In this article, we define inductive electron power deposition in the plasma as:

$$P_{e,\mathrm{ICP}} = J_{ze}E_{z(sol)}, \tag{2}$$

where $J_{ze}$ is the $z$-directed current density in the plasma.

The Ar plasma chemical mechanism in the Monte Carlo collision model considers electron–neutral elastic, inelastic, and ionization collisions. In addition, ion–neutral charge-exchange collisions are included. Surface interactions, including particle absorption and secondary processes, are treated through boundary models consistent with the electrostatic formulation.

## 3. Results

Two-dimensional (2D) inductively coupled plasma (ICP) simulations were performed in the geometry shown in Fig. 1. The simulations are done in Cartesian coordinates (*xy*) with a symmetry boundary condition at $x = 0$. The model includes four current-carrying coils embedded within a dielectric region (relative permittivity $\epsilon_r = 1$), all driven with identical currents flowing in the $z$-direction. A Faraday shield between the coils and the plasma is assumed (but not explicitly modeled) to suppress capacitive coupling from the voltage on the coils. Since capacitive coupling weakens at lower frequencies in ICP systems, simulations employing non-sinusoidal currents were conducted at 1 MHz. Non-inclusion of capacitive coupling between the coils and plasma is a limitation of the current modeling study. The plasma region has a half-width of $x = 15.06$ cm and a height of $y = 7.48$ cm. All simulation results reported in this article are for 25 mTorr gas pressure with gas temperature of 300 K. These simulations were run for 500 μs, which is sufficient time that important plasma properties reached steady state conditions for all conditions examined. The time-averaged results in this article correspond to average over the last 100 μs of the simulation.

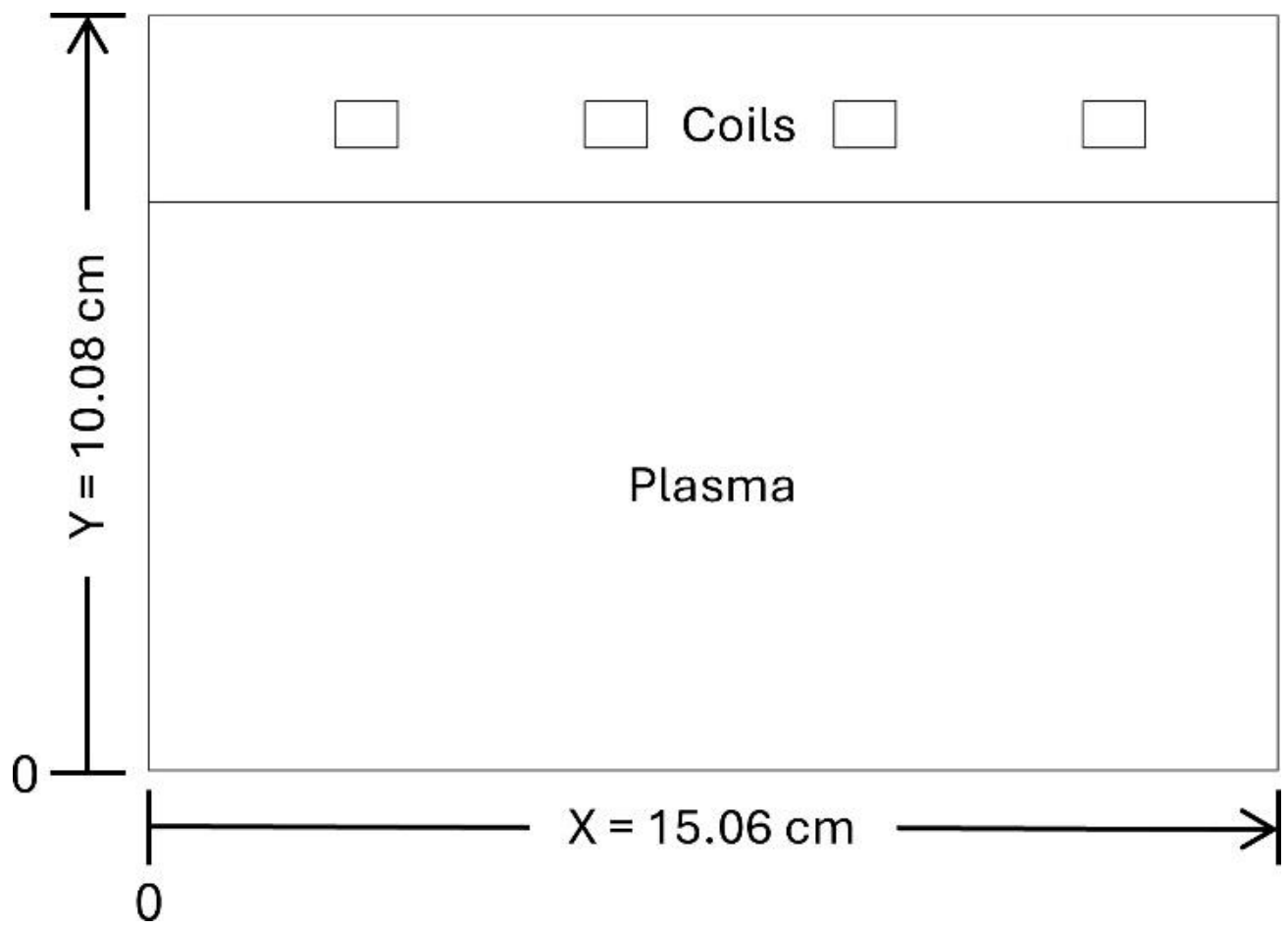

Fig. 1: Geometry used for the inductively coupled plasma simulations.

The current on the coils was either sinusoidal at different frequencies or asymmetric triangular waveform at 1 MHz, which is shown in Fig. 2(a). In both cases, the current amplitude in the article refers to the peak value. The duty cycle (DC) of the asymmetric triangular waveform is defined as

$$\mathrm{DC} = \frac{\tau_{\mathrm{up}}}{\tau_{\mathrm{up}}+\tau_{\mathrm{down}}}, \quad (3)$$

where $\tau_{\mathrm{up}}$ and $\tau_{\mathrm{down}}$ denote the linear ramp-up and ramp-down times, respectively. The harmonic content of these waveforms is plotted in Fig. 2(b). At DC = 70%, the first harmonic dominates, the second harmonic is noticeable, and higher harmonics are small. As DC increases, the first harmonic decreases while higher harmonics become increasingly more important. For DC > 90%, harmonics beyond the 10th are non-negligible, but they are not shown in Fig. 2(b).

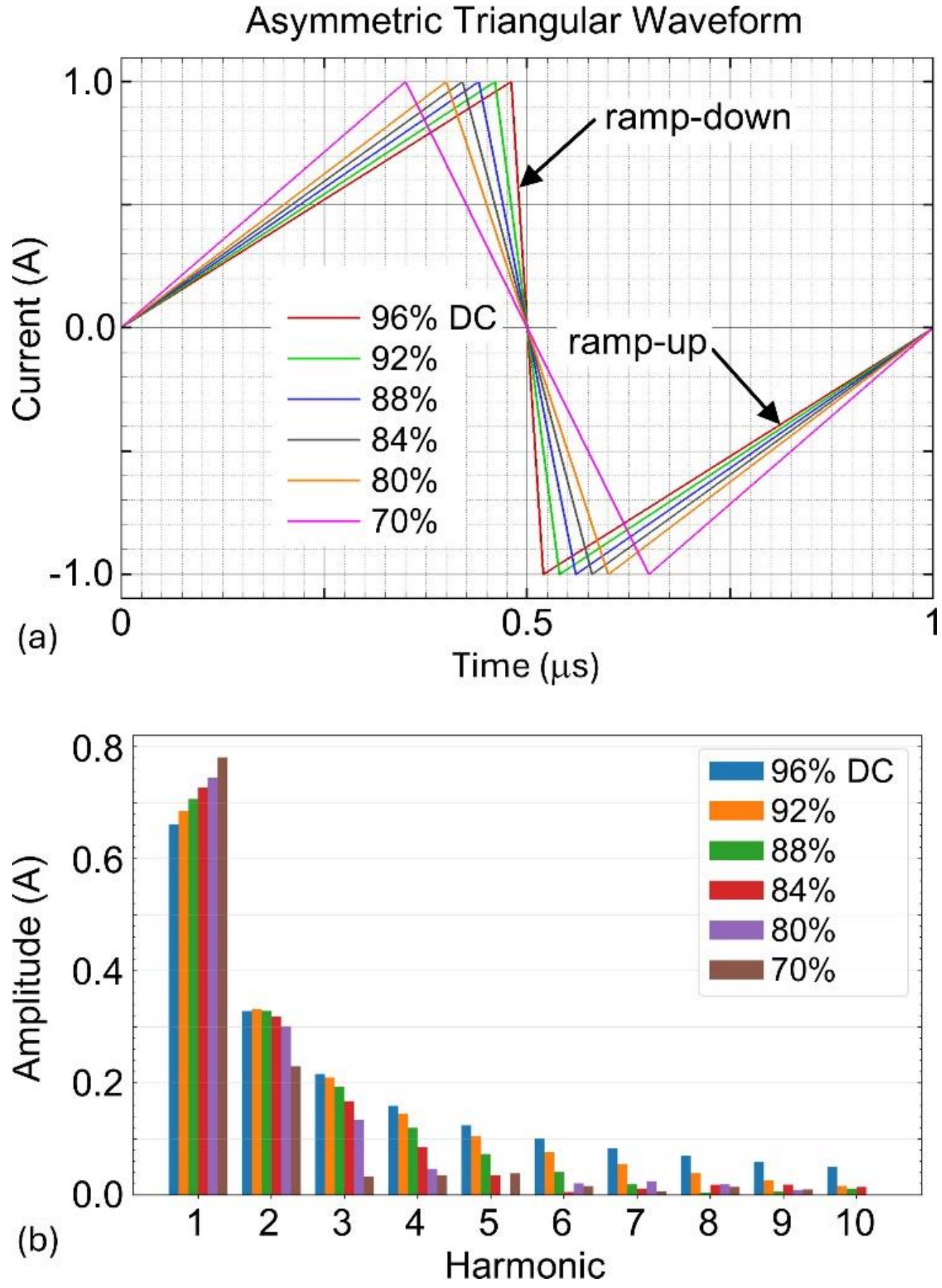

Fig. 2: (a) Asymmetric triangular current waveforms for duty cycles (DC) between 70 – 96%. (b) Harmonics of the asymmetric triangular waveform currents for DC between 70 – 96%

### 3.1 Sinusoidal Current Excitation

We first consider simulations with sinusoidal coil current. Some results for 1 MHz current with amplitude A are plotted in Fig. 3. These results include the time-averaged magnitude of the inductive electric field , the time-averaged electron power deposition $P_{e(ICP)}$, and the time-averaged electron temperature . The results in Fig. 3 correspond to averages over the last 100 RF cycles (100 μs) of the simulation. Due to constructive interference, the electric field is strongest beneath the two inner coils. It decays vertically within the plasma. Owing to the small chamber dimensions, the field extends throughout the plasma volume and reaches the chamber boundaries. For this baseline case, peak $s^{-1}$. As , the plasma is collisional and the collisional skin depth cm. [31] As shown in Fig. 3(b), power deposition peaks between the central coils in the -direction and extends over a finite distance in the -direction, consistent with the electron density maximum located near the chamber center (Fig. 4). This inductive heating sustains the discharge through ionization collisions. The steady-state electron temperature $T_e$ profile, shown in Fig. 3(c), peaks slightly below the region of maximum power deposition and is relative uniform in the plasma region.

Figure 4 presents steady state electron density ($n_e$) profiles for frequency (*f*) in the range 1 – 30 MHz. At each frequency, $n_e$ is found to increase with increasing coil current amplitude. As the inductive electric field scales as $f I_{\text{coil}}$ in Eq. 1, significantly lower coil current is required at higher frequencies to achieve comparable peak plasma density. For the results plotted in Fig. 4, the coil currents are chosen to yield peak $n_e$ values within a similar range, although the resulting peak and volume-averaged densities are not identical. While the overall density profiles differ slightly with frequency, strong diffusion at 25 mTorr in the small chamber leads to broadly similar spatial distributions. With increasing frequency, the region of peak $n_e$ shifts closer to the plasma–dielectric interface and outwards horizontally.
The RF coil currents launch an electromagnetic wave, and the electrons accelerate in the resulting electric field. These electrons undergo collisions leading to ionization and excitation. As the skin depth decreases with increasing frequency, [31] power deposition becomes more localized near the plasma–dielectric interface at higher frequency. This trend is illustrated in Figs. 5 and 6, where we have plotted the argon excitation () and ionization () collision frequencies, respectively. Although peak collision frequencies remain comparable across frequencies - owing to the adjusted coil currents - the location of peak collisional activity clearly moves toward the interface as the frequency increases.

The above results for the effect of frequency are not novel and are consistent with our current understanding of ICPs. They have been included in this paper to interpret the results for non-sinusoidal current waveform, which are discussed next.

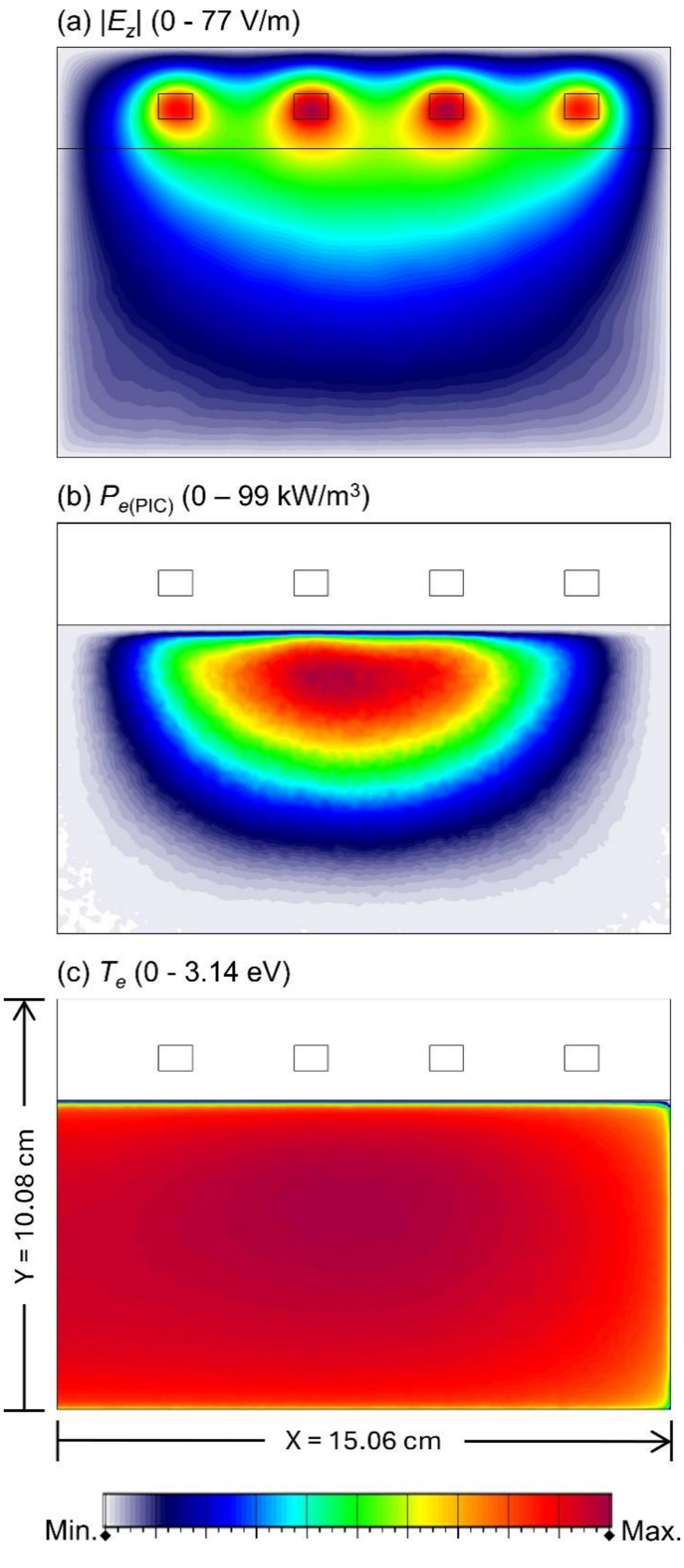

Fig. 3: (a) Amplitude of the inductive electric field $|E_z|$, (b) inductive power deposition $P_{e(ICP)}$, and (c) electron temperature $T_e$. These results are for Ar plasma at 25 mTorr produced by 35 A sinusoidal current in all coils at 1 MHz.

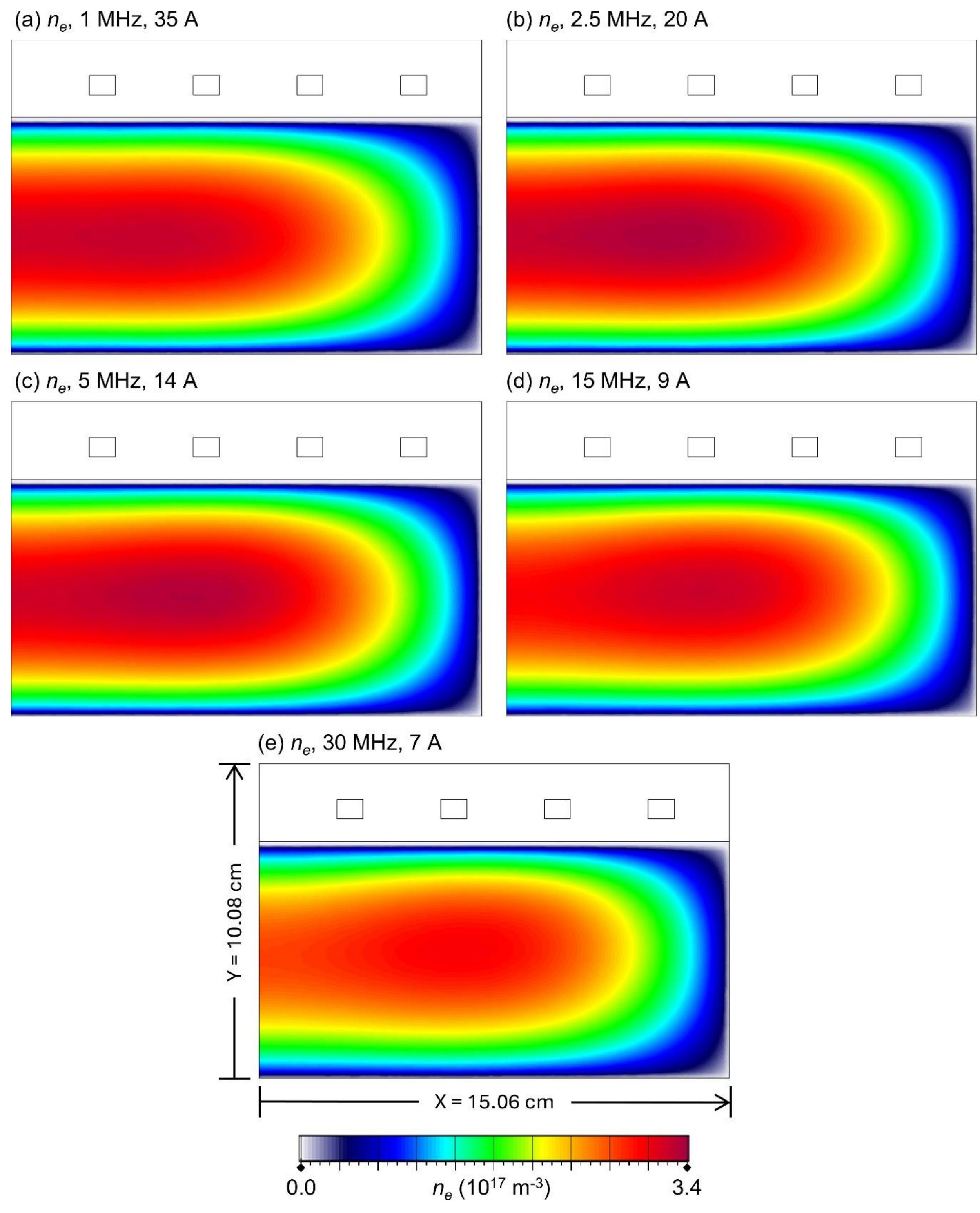

Fig. 4: Time-averaged electron density $n_e$ for $I_{coil}$ = (a) 35 A at 1 MHz, (b) 20 A at 2.5 MHz, (c) 14 A at 5 MHz, (d) 9 A at 15 MHz, and (e) 7 A at 30 MHz. These results are for sinusoidal coil current, and Ar plasma at 25 mTorr.

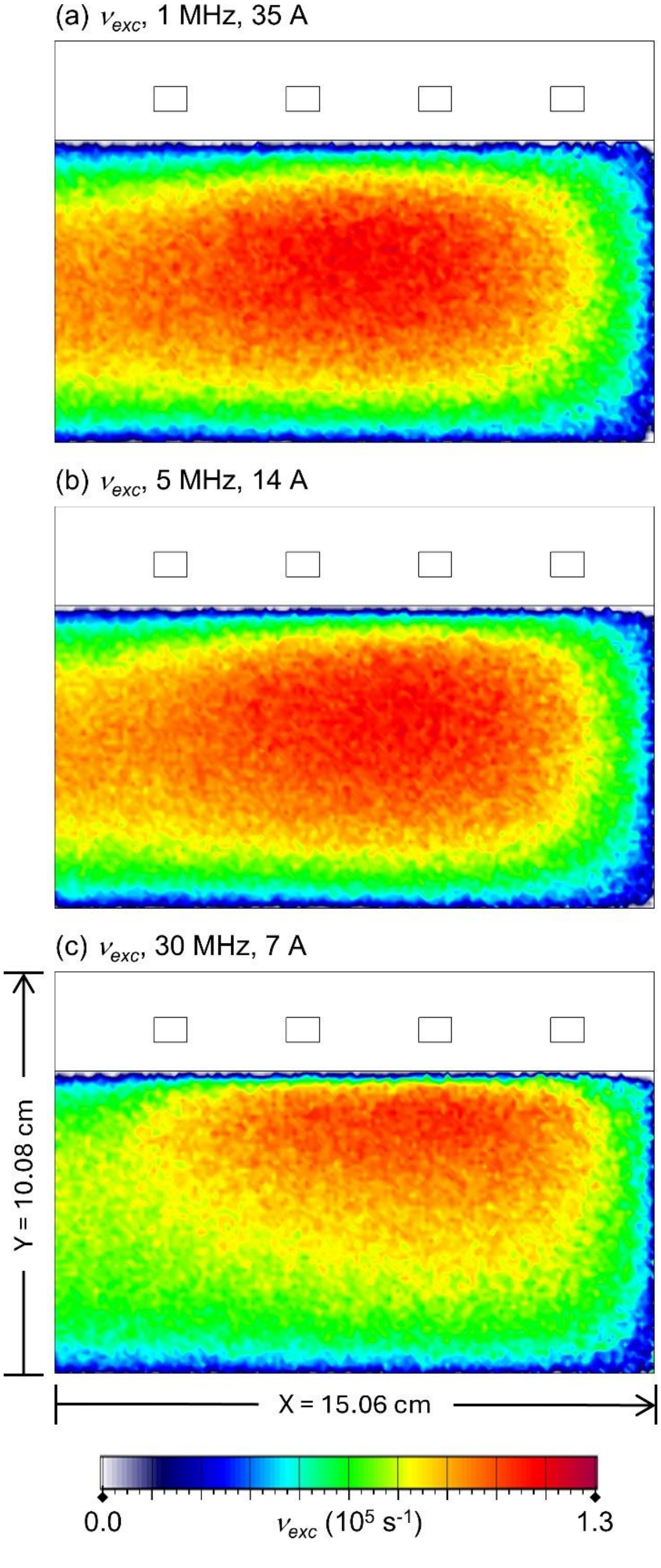

Fig. 5: Time-averaged collision frequency for the metastable excitation collision $\nu_{exc}$ for $I_{coil}$ = (a) 35 A at 1 MHz, (b) 14 A at 5 MHz, and (c) 7 A at 30 MHz. These results are for sinusoidal coil current, and Ar plasma at 25 mTorr.

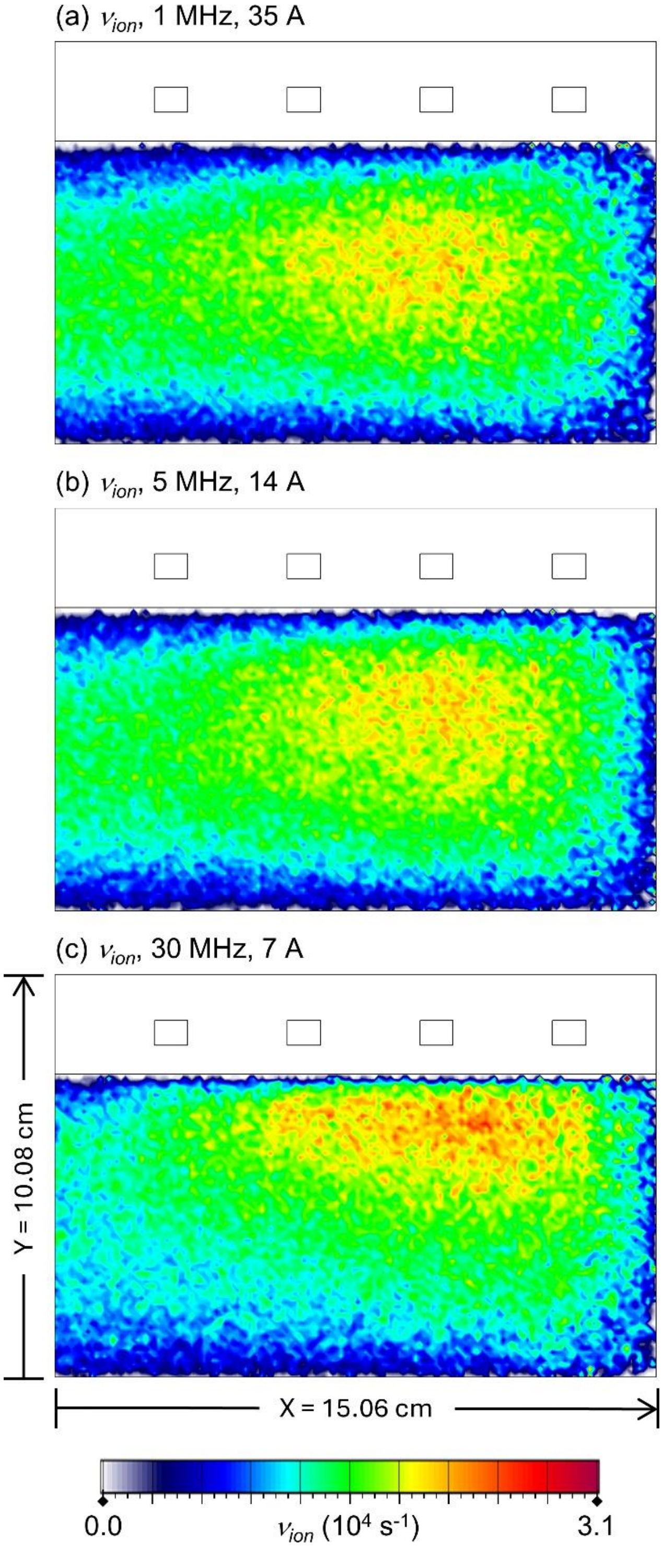

Fig. 6: Time-averaged collision frequency for the ionization collision $\nu_{ion}$ for $I_{coil}$ = (a) 35 A at 1 MHz, (b) 14 A at 5 MHz, and (c) 7 A at 30 MHz. These results are for sinusoidal coil current, and Ar plasma at 25 mTorr.

### 3.2 Asymmetric Triangular Waveform Excitation

In this section, we examine plasmas driven by asymmetric triangular coil current. Because the inductive electric field scales with $\partial J_{\text{ext}}/\,\partial t$, and $\tau_{\text{down}} < \tau_{\text{up}}$ for the waveforms in Fig. 2, the electric field is strongest during the current ramp-down phase. Plasma production therefore increases during this phase. However, since the steady-state plasma density is attained over many RF cycles, the net change in plasma density within a single cycle is small. Figure 7 shows the time-averaged electron density $n_e$ for DC values between 70% and 96%, obtained with a fixed coil current amplitude of $I_{\text{coil}} = 35$ A. The relative amplitude of higher current harmonics increases with DC (Fig. 2(b)), leading to higher steady-state $n_e$ at higher DC. Due to strong diffusion at 25 mTorr, the overall density profiles remain insensitive to DC, although the density peak weakly shifts toward larger $x$ and higher $z$ with increasing DC.

Figures 8 and 9 present the time-averaged argon excitation ($\nu_{\text{exc}}$) and ionization ($\nu_{\text{ion}}$) collision frequencies as a function of DC. Like $n_e$, peak collision frequencies increase with DC. In contrast to the density profiles, however, the influence of the waveform harmonic content is clearer in the collision distributions: increasing DC shifts the dominant collisional activity closer to the plasma–dielectric interface.

Although both the time-averaged electron-impact excitation and ionization collision frequencies increase with DC, their relative rates of increase differ due to changes to the electron energy distribution function (EEDF). The ratio of volume and time-averaged collision frequencies, $\nu_{\text{exc}}/\nu_{\text{ion}}$, is plotted in Fig. 10, and indicates that excitation collisions become more dominant at higher DC. To elucidate this trend, we have plotted spatially averaged $\nu_{\text{exc}}$ and $\nu_{\text{ion}}$ as a function of time for different DC values in Fig. 11. Because the inductive electric field and power deposition peak during the current ramp-down phase, both collision frequencies rise rapidly during this interval. As discussed later, the population of high energy electrons increases during this phase. Consequently, the frequency of ionization collisions, which have a higher threshold energy, increases more rapidly at higher DC. However, $\nu_{\text{ion}}$ also decreases more during the current ramp-up phase at higher DC. As $\tau_{\mathrm{up}} < \tau_{\mathrm{down}}$ for the asymmetric triangular waveforms considered here, the ramp-up phase has a bigger impact on the time-averaged $\nu_{\mathrm{exc}}/\nu_{\mathrm{ion}}$. As a result, $\nu_{\mathrm{exc}}/\nu_{\mathrm{ion}}$ increases with DC (Fig. 10).

### 3.3 Electron Velocity Distribution Function and Time-Resolved Behavior

One major objective of this study is to assess how current waveform asymmetry modifies the EEDF or electron velocity distribution function (EVDF). Figure 12 shows the time-averaged EVDF for several DC values. These EVDFs have been averaged over the region $x = 4.99$–10.62 cm and $y = 3.79$–7.58 cm, which corresponds to the region of intense inductive power deposition. The electron velocity distributions have been plotted vs. $m_e v_{ze}|v_{ze}|/(2e)$, which preserves the electron directional information. Because electrons experience stronger acceleration during the ramp-down phase, the EVDF becomes increasingly asymmetric as DC increases. To elucidate the underlying mechanisms behind the time-dependent collision frequency data in Fig. 11, Fig. 13 presents time-dependent EVDFs for DC = 96%, averaged over the same region as Fig. 12. The EVDF is symmetric at 0.42 μs, well before the onset of current ramp-down. At 0.50 μs, near the midpoint of the ramp-down phase, the enhanced inductive electric field increases the high-energy electron population, thereby strongly enhancing ionization and excitation, particularly the former due to its higher energy threshold. Although the electric field primarily accelerates electrons in the $-z$ direction, collisions redistribute energy, increasing the population of electrons traveling in the $+z$ direction as well. After the current begins ramping up at 0.52 μs, more than 0.1 μs is required for the EVDF to relax to its pre-ramp-down state.

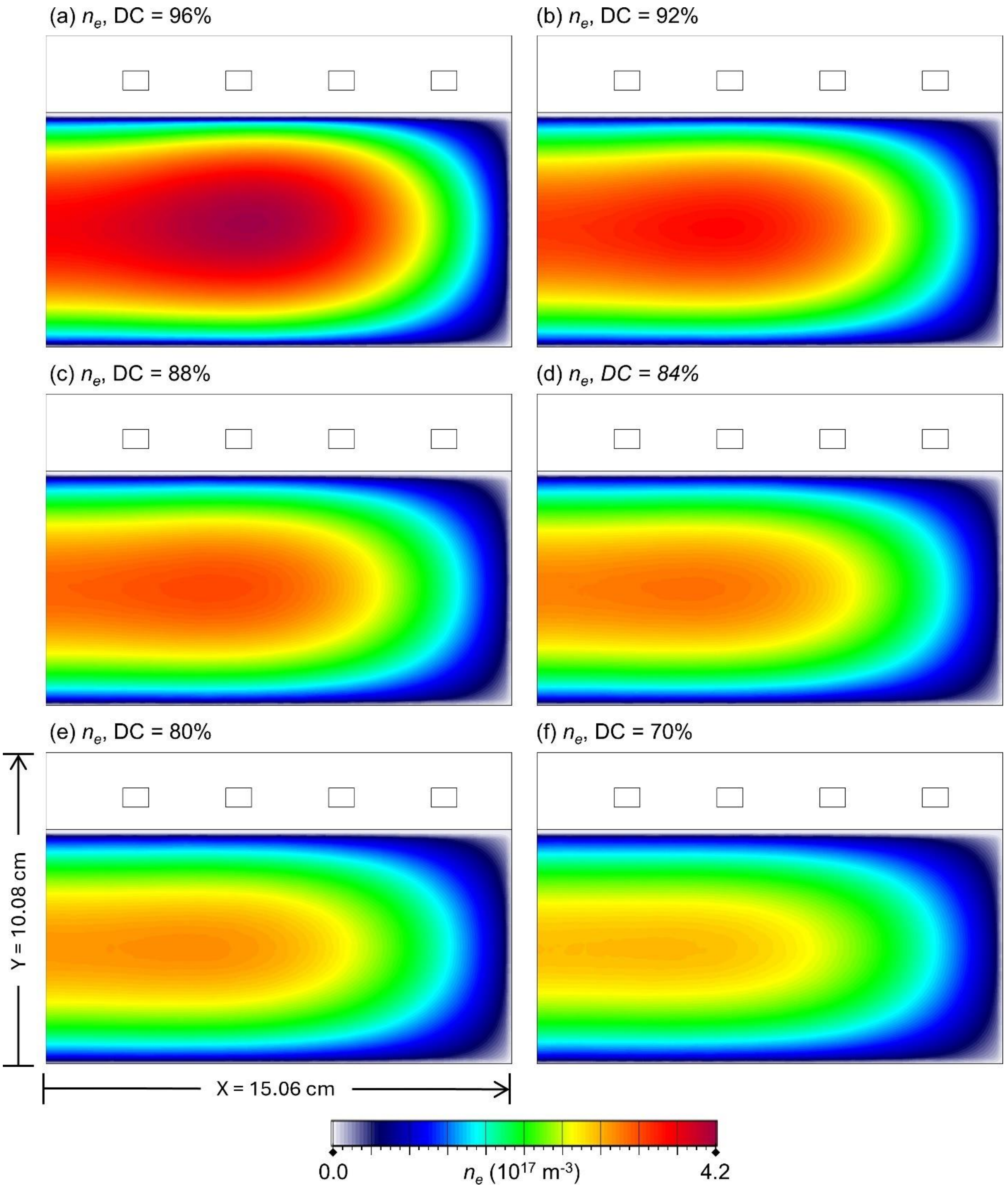

Fig. 7: Time-averaged electron density $n_e$ for $I_{coil}$ = 35 A at 1 MHz. These results are for asymmetric triangular waveform with DC = (a) 96%, (b) 92%, (c) 88%, (d) 84%, (e) 80%, and (f) 70%. The gas pressure in Ar plasma is 25 mTorr.

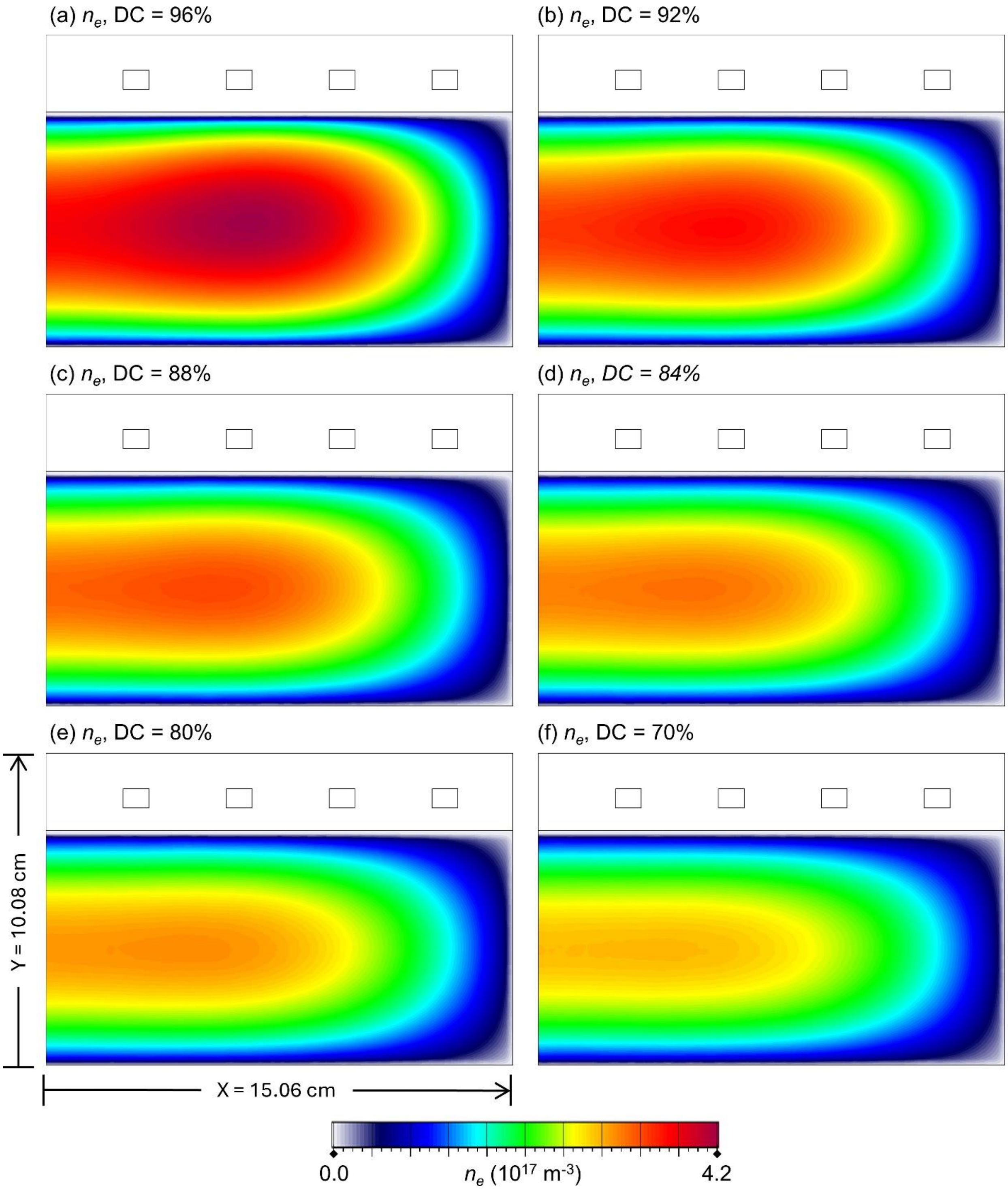

Fig. 8: Time-averaged metastable Ar excitation collision frequency $\nu_{exc}$ for $I_{coil}$ = 35 A at 1 MHz. These results are for asymmetric triangular waveform with DC = (a) 96%, (b) 92%, (c) 88%, (d) 84%, (e) 80%, and (f) 70%. The gas pressure in Ar plasma is 25 mTorr.

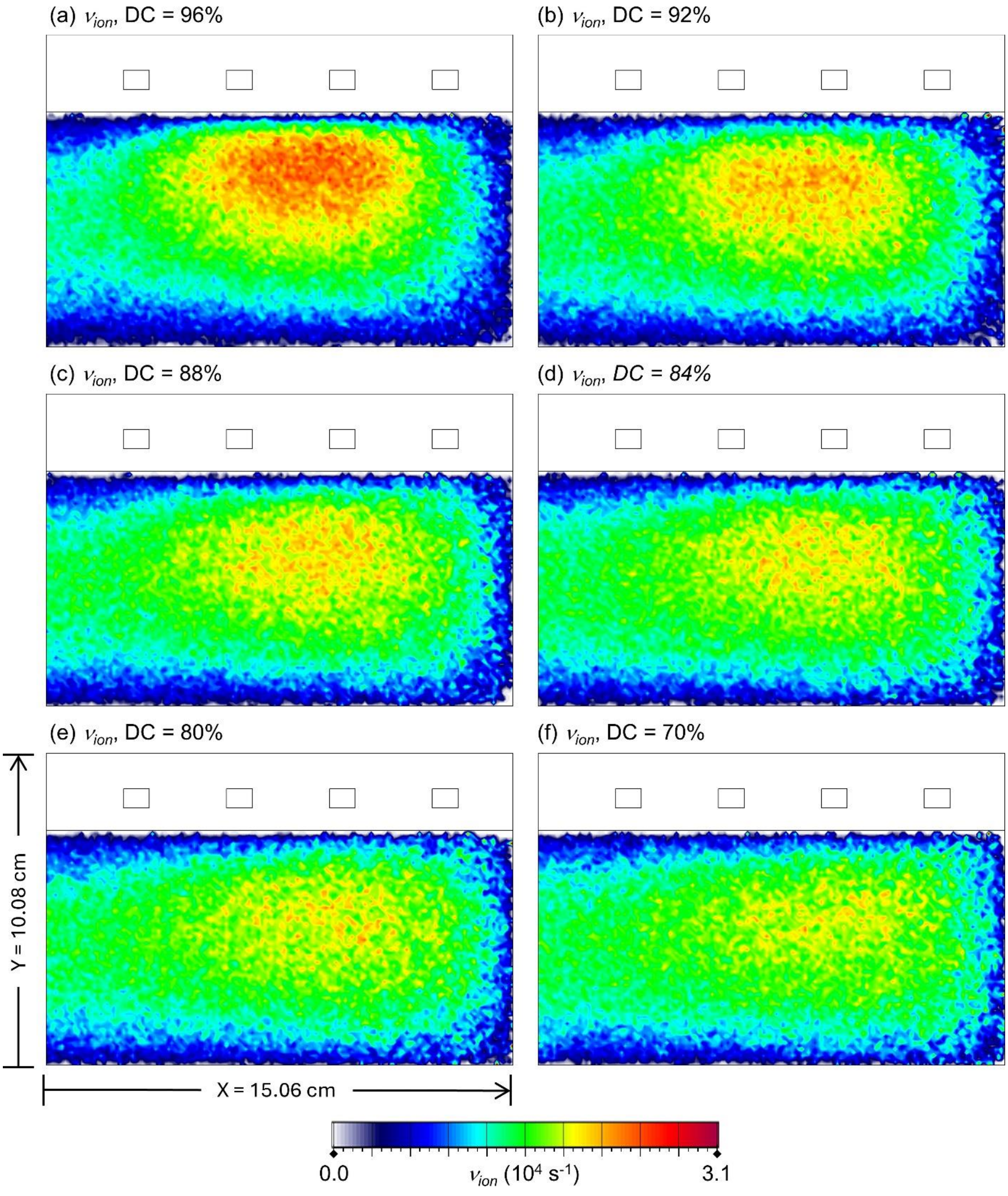

Fig. 9: Time-averaged Ar ionization collision frequency $\nu_{ion}$ for $I_{coil}$ = 35 A at 1 MHz. These results are for asymmetric triangular waveform with DC = (a) 96%, (b) 92%, (c) 88%, (d) 84%, (e) 80%, and (f) 70%. The gas pressure in Ar plasma is 25 mTorr.

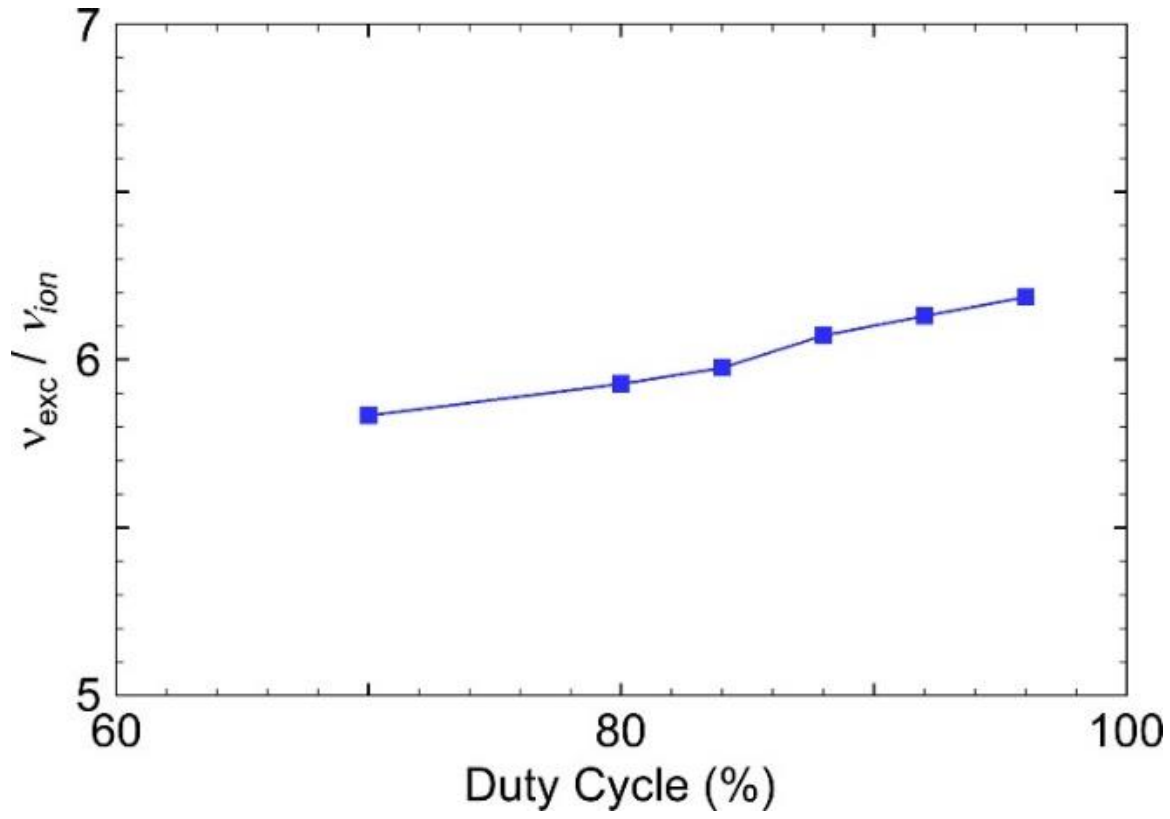

Fig. 10: Volume and time-averaged ratio of $\nu_{\mathrm{exc}}/\nu_{\mathrm{ion}}$ vs. duty cycle for asymmetric triangular waveform currents. These results are for $I_{coil} = 35$ A at 1 MHz. The gas pressure in Ar plasma is 25 mTorr.

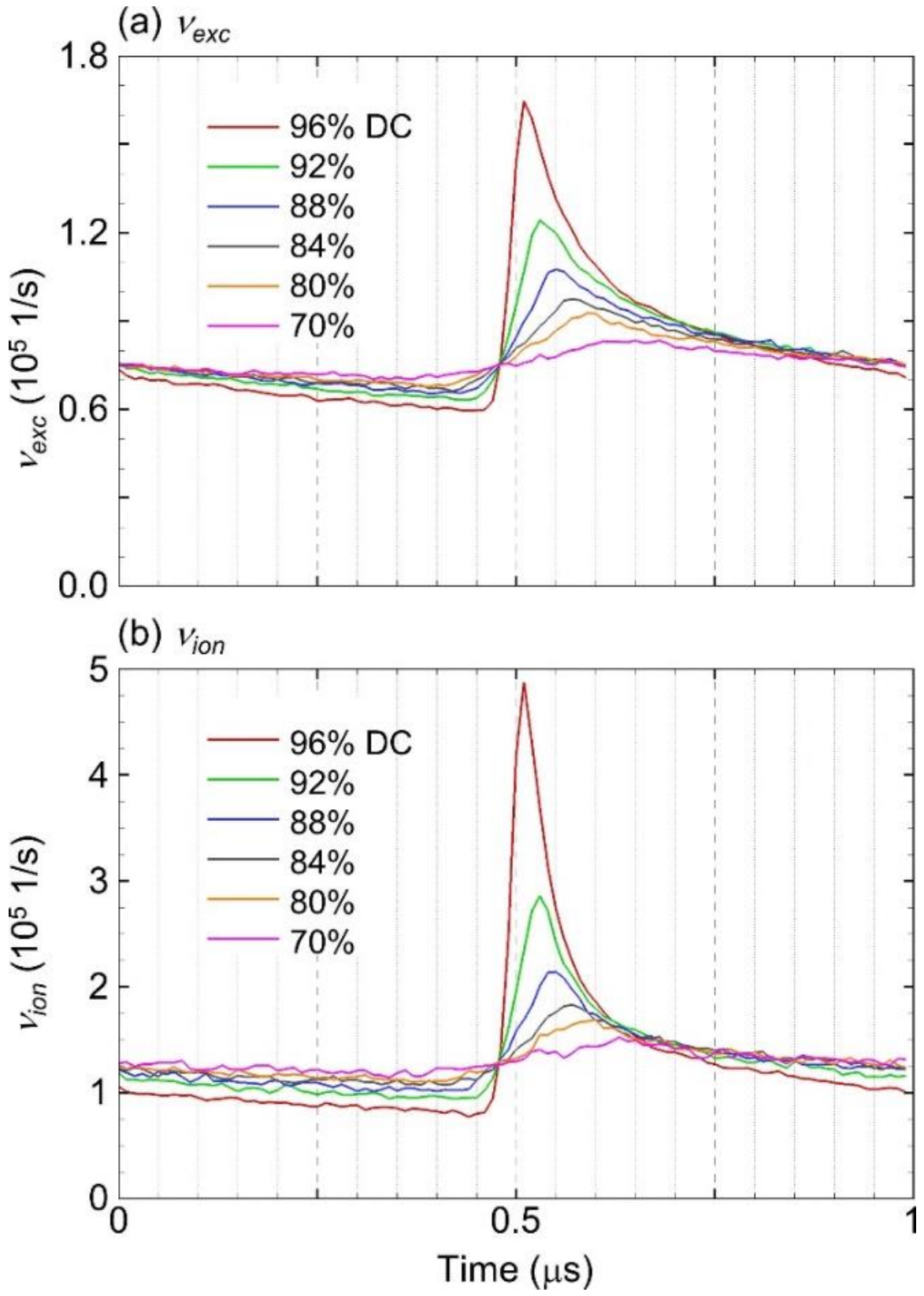

Fig. 11: Volume -averaged collision frequencies (a) $\nu_{\mathrm{exc}}$ and (b) $\nu_{\mathrm{ion}}$ plotted as a function for time for several duty cycles. These results are for asymmetric triangular waveform currents with $I_{coil} = 35$ A at 1 MHz. The gas pressure in Ar plasma is 25 mTorr.

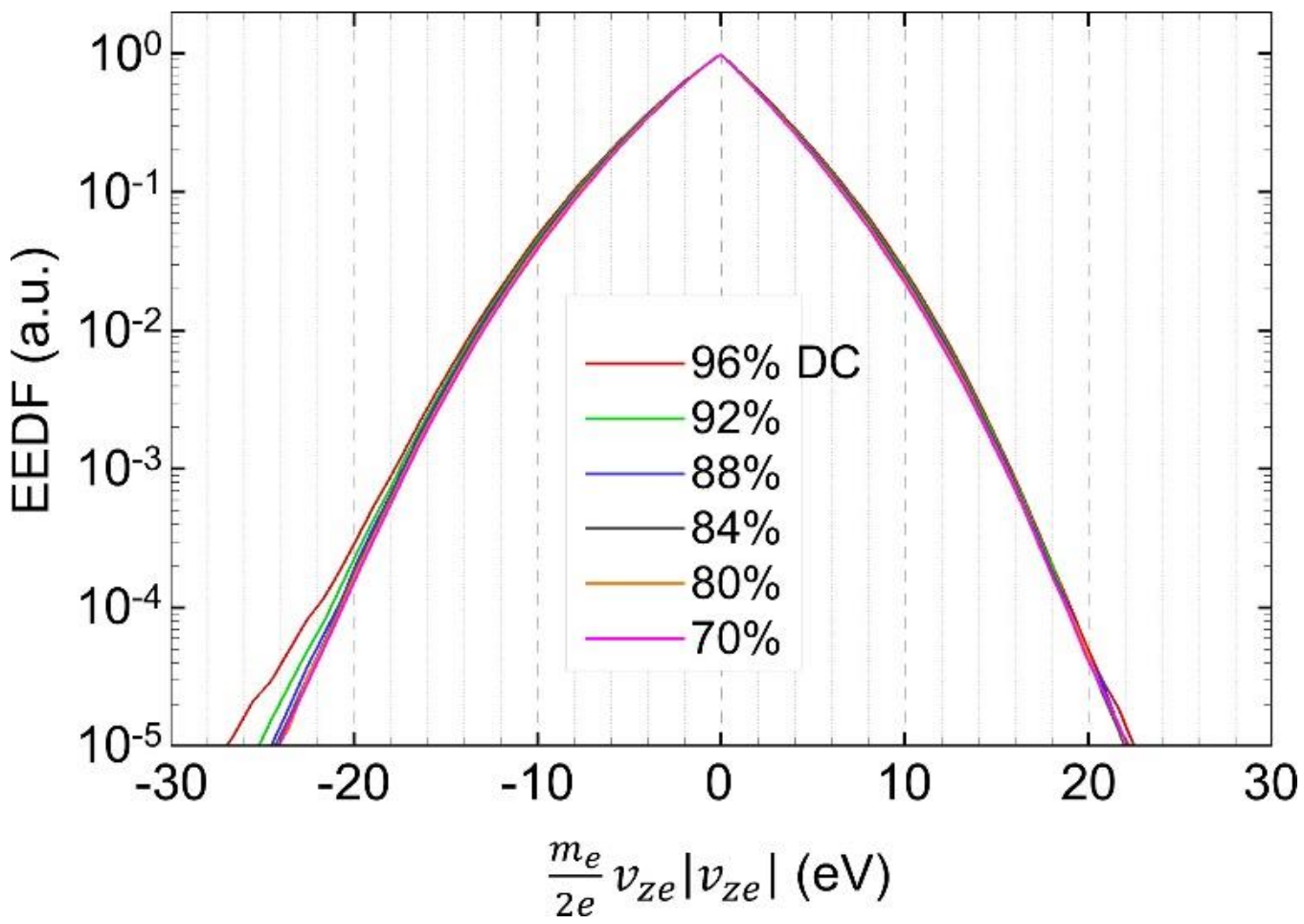

Fig. 12: The time-averaged electron velocity distribution function (EVDF) for different duty cycles. The EVDF has been averaged over $x = 4.99$–$10.62$ cm and $z = 3.79$–$7.58$ cm. These results are for asymmetric triangular waveform currents with $I_{coil} = 35$ A at 1 MHz. The gas pressure in Ar plasma is 25 mTorr.

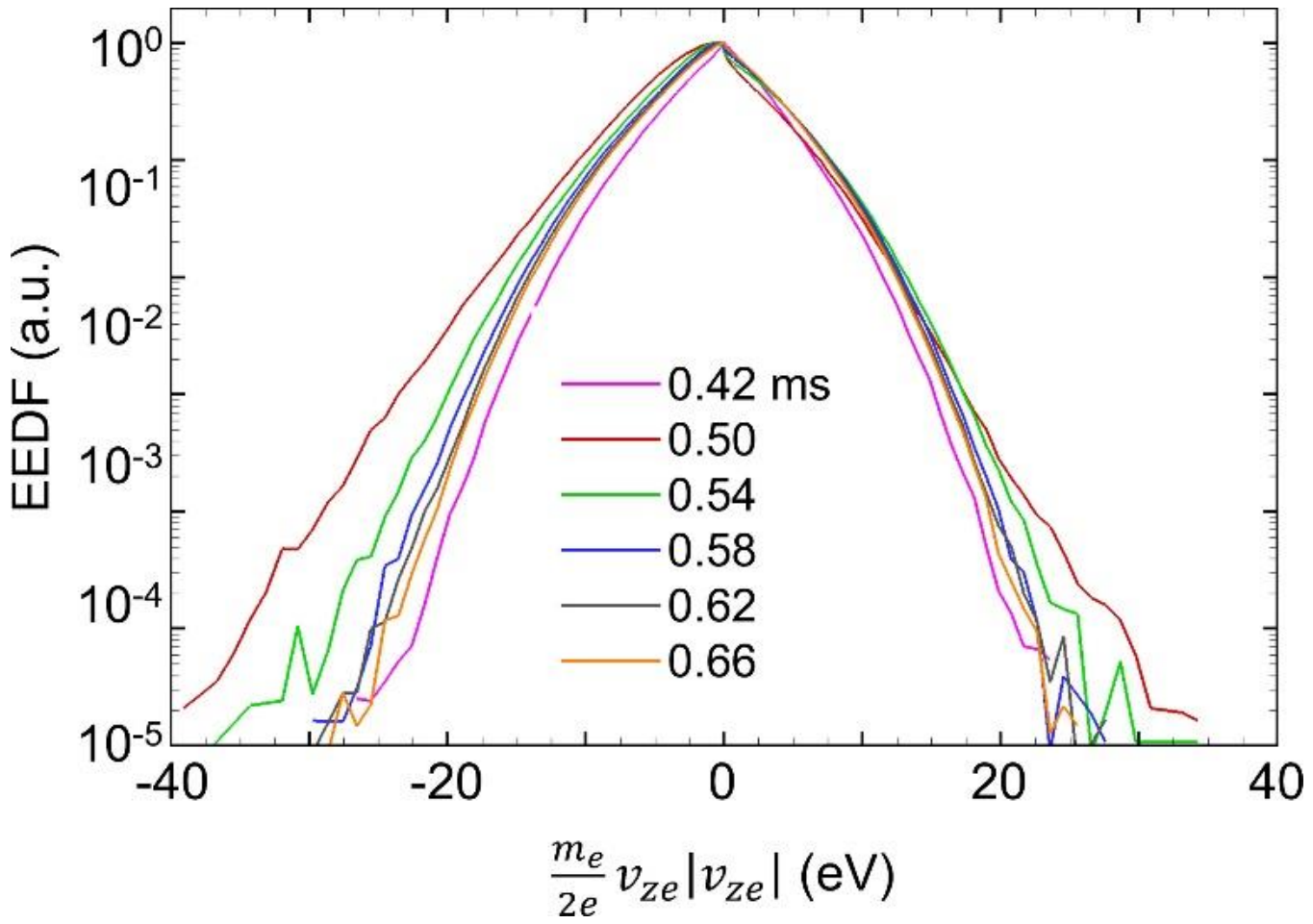

Fig. 13: The time-dependent electron velocity distribution function (EVDF) for DC = 96% at several times. The EVDF has been averaged over $x = 4.99$–$10.62$ cm and $z = 3.79$–$7.58$ cm. These results are for asymmetric triangular waveform currents with $I_{coil} = 35$ A at 1 MHz. The gas pressure in Ar plasma is 25 mTorr.

The rapid change in coil current during the ramp-down phase introduces transient effects in the plasma that persists for many 10s of ns. Figures 14 and 15 show the temporal evolution of $E_z$, $J_{ze}$, and $P_{e,\mathrm{ICP}}$ in two planes, $x = 7.53$ cm and $z = 5.04$ cm, respectively, for DC = 96%. All quantities peak during the current ramp-down phase, with electron current flow and power deposition occurring primarily below the plasma–dielectric interface. Although the source-region electric field reverses direction after 0.52 μs, the electric field within the plasma remains positive for more than 0.1 μs due to the inertia of the plasma current. The current in the plasma induces an electric field and electrons continue to absorb power during this time.

## 4. Summary

Two-dimensional particle-in-cell simulations were performed to investigate the effects of driving frequency and current waveform on plasma characteristics in an inductively coupled plasma (ICP) operating at 25 mTorr. For simulations with sinusoidal excitation, we attempted to obtain comparable peak electron densities. The required coil current to maintain similar plasma density decreased monotonically with the driving frequency from 1 to 30 MHz, consistent with the scaling of the inductive electric field with frequency. As frequency increased, power deposition and the rate of inelastic collisions shifted closer to the plasma – dielectric interface due to reduced skin depth. Despite these changes, strong diffusion in the small chamber resulted in broadly similar electron density profiles across the frequency range considered.

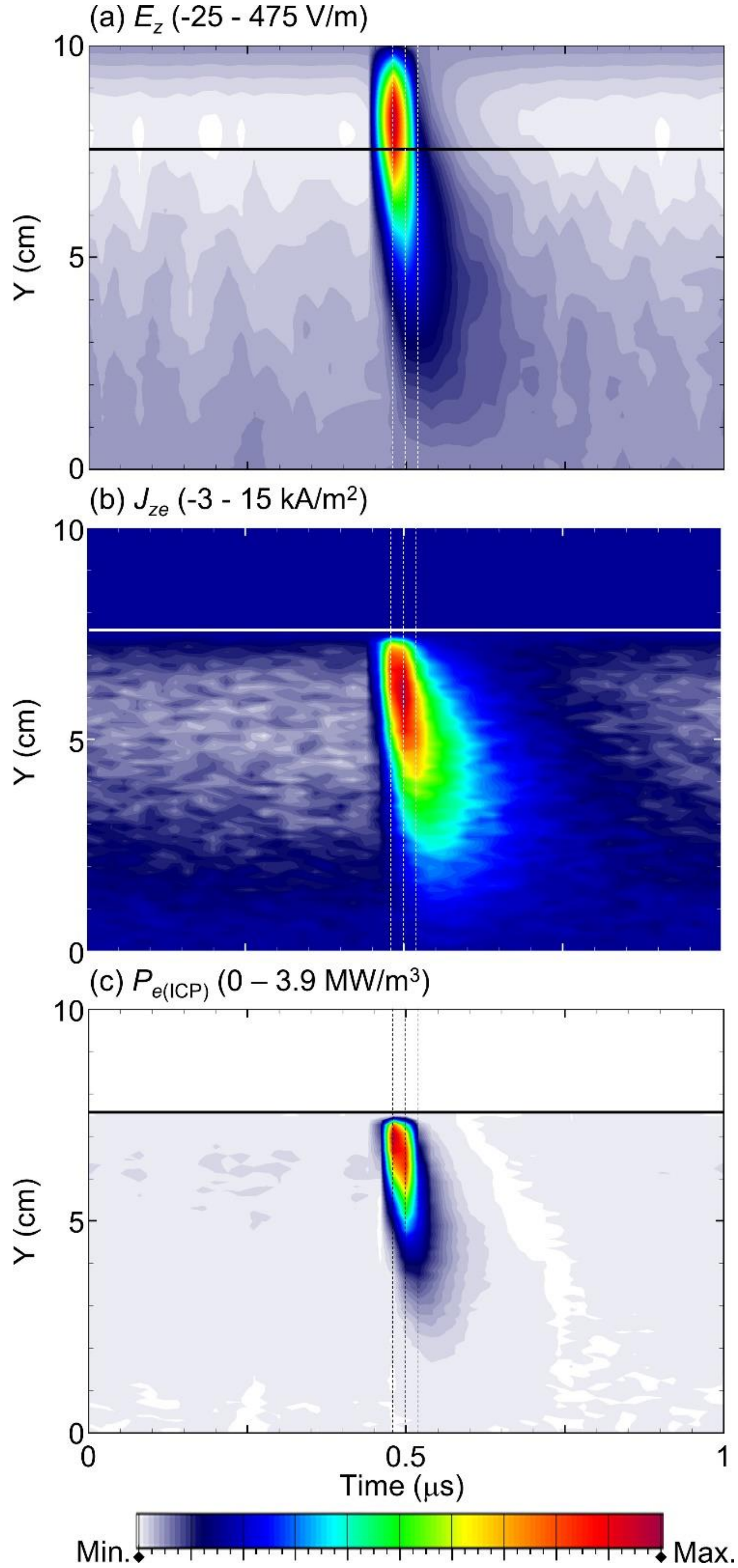

Fig. 14: Time resolved (a) $E_z$, (b) $J_{ze}$, and (c) $P_{e(ICP)}$ in the plane $x = 7.53$ cm for DC = 96%. These results are for asymmetric triangular waveform currents with $I_{coil} = 35$ A at 1 MHz. The gas pressure in Ar plasma is 25 mTorr.

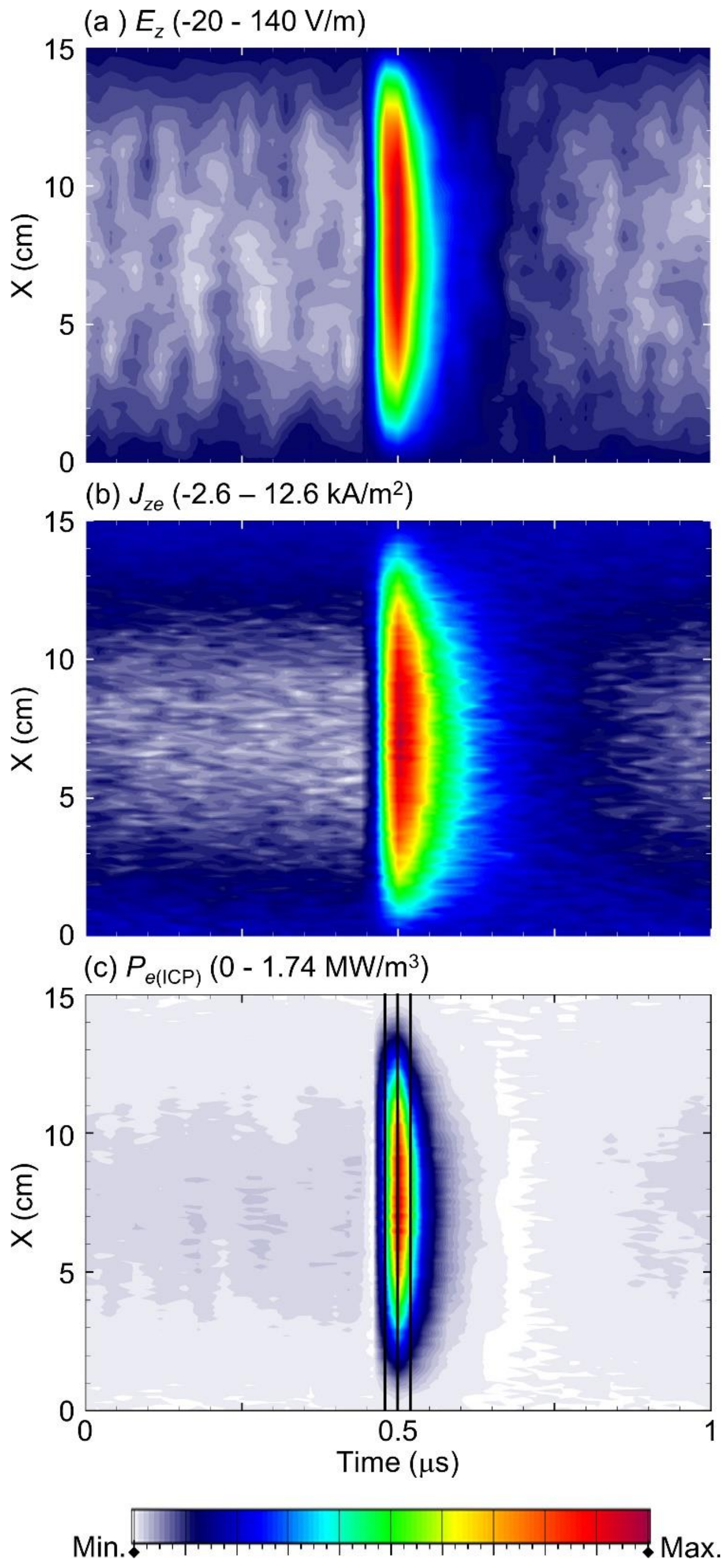

Fig. 15: Time resolved (a) $E_z$, (b) $J_{ze}$, and (c) $P_{\mathrm{e(ICP)}}$ in the plane $z = 5.04$ cm for DC = 96%. These results are for asymmetric triangular waveform currents with $I_{coil} = 35$ A at 1 MHz. The gas pressure in Ar plasma is 25 mTorr.

The impact of current waveform asymmetry was examined for asymmetric triangular waveforms at 1 MHz. Because the inductive electric field scales with the time derivative of the coil current, electron heating and plasma production were strongly enhanced during the current ramp-down phase. The contribution of higher harmonics increases with duty cycle (DC). This led to higher steady-state electron density and excitation and ionization processes occurring closer to the plasma – dielectric interface at higher DC. While the overall density profiles were only weakly affected by DC, the spatial distributions of collisional processes exhibited clear sensitivity to waveform asymmetry.

The electron velocity distribution function (EVDF) was found to be strongly influenced by the asymmetric current waveform. Increasing DC led to more asymmetric EVDFs, reflecting preferential electron acceleration during the current ramp-down phase. Time-resolved analysis showed that excitation and ionization rates peak sharply during this phase, with ionization exhibiting stronger relative enhancement due to its higher energy threshold. Due to the strong current produced in the plasma during the ramp-down phase, transient plasma currents persisted well beyond the reversal of the source electric field. The self-induced electric led to continued electron power deposition in the plasma for 10s of ns after the coil current reverses direction.

Overall, these results demonstrate that current waveform shape provides an effective means of controlling electron power deposition, EEDF, EVDF, and electron-impact collisional processes in ICPs. Current waveform tailoring offers a promising approach to control the EEDF and selectively influence electron-impact collision dynamics without substantial changes to the global plasma density.

**Acknowledgements**

Shahid Rauf would like to acknowledge useful discussions with Julian Schulze, Ya Zhang, and Wei Jiang.

**Author Declarations**

**Conflict of Interest**

The authors have no conflicts to disclose.

**Data Availability Statement**

The data that supports the findings of this study are available from the corresponding author upon reasonable request.